%% file: elsarticle-template-harv.tex
\documentclass[preprint,review,3p,times,authoryear]{elsarticle}
\usepackage{lineno}
\usepackage{subfig}
\journal{JMPS}

\input{mypreamble}
\input{symbolcommands}
\begin{document}

\begin{frontmatter}


\author[fzj,tubaf]{M. Budnitzki}
\ead{m.budnitzki@fz-juelich.de}
\author[fzj,tubaf]{S. Sandfeld}
\ead{s.sandfeld@fz-juelich.de}
\address[fzj]{Institute for Advanced Simulation (IAS-9: Materials Data Science and Informatics), Forschungszentrum Jülich GmbH, 52428 Jülich, Germany}
\address[tubaf]{TU Bergakademie Freiberg, Institute of Mechanics and Fluid Dynamics, Lampadiusstr.~4, 09599 Freiberg}

\title{A model for the interaction of dislocations with planar defects based on Allen-Cahn type microstructure evolution coupled to strain gradient elasticity}

\begin{abstract}
In classical elasticity theory the stress-field of a dislocation is characterized by a $1/r$-type singularity. When such a dislocation is considered together with an Allen-Cahn-type phase-field description for microstructure evolution this leads to singular driving forces for the order parameter, resulting in non-physical (and discretization-dependent) predictions for the interaction between dislocations and phase-, twin- or grain-boundaries. We introduce a framework based on first strain gradient elasticity to regularize the dislocation core. It is shown that the use of strain energy density that is quadratic in the gradient of elastic deformation results in non-singular stresses but may result in singular driving forces, whereas a strain energy, which is quadratic in the gradient of the full deformation tensor, regularizes both stresses and driving forces for the order parameter and is therefore a suitable choice. The applicability of the framework is demonstrated using a comprehensive example.

\end{abstract}


\end{frontmatter}


\section{Introduction}
\label{sec:introduction}

Phase field approaches have proven to be very powerful for the investigation of the formation and evolution of microstructures due to solid-solid phase transformations and twinning. This appears to be the natural framework for the investigation of the interaction of planar crystal defects such as phase- or twin-boundaries with line defects (dislocations, disclinations). A typical phase field model for diffusionless (martensitic) transformations comprises of evolution equations of \textsc{Allen-Cahn}-type for the order parameters $\phi_{\beta}$
\begin{equation}
  \label{eq:AC}
  M^{-1} \sdot\phi_{\beta} = \alpha\Delta\phi_{\beta} - \rho\pder{\psi}{\phi_{\beta}}\,,
\end{equation}
where $M$ and $\alpha$ are constants, $\rho$ denotes the mass density, and $\psi$ is a bulk specific free energy. The subscript $\beta$ indicates the number of the phase, grain or twin variant. Assuming a small perturbation setting, the linear strain tensor\footnote{Nomenclature: We denote vectors by bold lower case latin $\vec{a}$ and greek $\vec{\alpha}$ letters. The dot operator ``$\cdot$'' denotes the scalar product.
Second order tensors are denoted by bold uppercase latin letters $\vec{A}$. We introduce a scalar product between second order tensors denoted by ``:'' as $\tA:\tB:=\tr{\tA\cdot\tB^{\top}}$, where $\tB^{\top}$ is the transpose of $\tB$ and $\tr{(\cdot)}$ denotes the trace operator. Similarly, we denote third order tensors $\tria{A}$ by bold calligraphic capital letters and ``$\smash\tridot$'' is the corresponding scalar product. We use black-board capital letters $\tet{C}$ for fourth-order tensors. Whenever index-notation is used, summation over latin indices appearing twice is implied and spatial derivatives are denoted using the comma operator, e.g. $\pder{y}{x_{i}} \equiv y^{,i}$.} $\tE$ can be additively decomposed into elastic $\tE^{\ue}$ and inelastic (i.e., eigenstrain) $\tEtr$ contributions, such that $\tEe = \tE - \tEtr$. Assuming linear elasticity, the stress $\tS$ is given by $\tS=\bbC:\tEe$, and  the specific free energy takes the form
\begin{equation}
  \label{eq:bulk_free_energy}
  \psi\bigl( \tE\,,\,\phi_{\beta}\,,\, \theta \bigr) = \half \tEe : \bbC : \tEe + \psi_{\ub}\bigl( \phi_{\beta}\,,\, \theta \bigr)\,.
\end{equation}
As a consequence, the evolution equation~\eqref{eq:AC} can be rewritten as
\begin{equation}
  \label{eq:AC_exp}
  M^{-1}\sdot\phi_{\beta} = \alpha\Delta\phi_{\beta} + \tS:\pder{\tEtr}{\phi_{\beta}} -  \rho\pder{\psi_{\ub}}{\phi_{\beta}}\,.
\end{equation}
In linear elastic Volterra theory, the stresses diverge as the dislocation line is approached. In particular for dislocations the singularity is of $1/r$-type. As per Eq.~\eqref{eq:AC_exp}, this results in singular driving forces for the evolution of the order parameters, effectively negating the concepts such as a nucleation barrier or a pile-up stress. Different approaches to regularize the stress in the core region exist in literature based either on the concept of a distributed \textsc{Burger}'s vector \citep{Lothe:1992,Cai:2006}, which are inspired by richer microscopic models for dislocations \citep{Peierls:1940,Nabarro:1947}, or generalized continuum theories \citep{Lazar:2005,Lazar:2006,Lazar:2015,Po:2018}. However, the first strain gradient approach advocated by \citet{Po:2018} has the advantage that the obtained regularization is independent of the type of defect in question and therefore does not require any defect-specific information for the determination of model parameters. In principle, these parameters can directly be obtained from atomistic interaction potentials \citep{Admal:2017}.

The purpose of this work is to follow a micromorphic approach and to derive a framework which consistently couples first strain gradient elasticity to \textsc{Allen-Cahn}-type microstructure evolution ensuring non-singular driving forces on the order parameters in the presence of line defects.

\section{Balance equations and boundary conditions}
\label{sec:therm-coupl}

The principle of virtual power (PVP) provides a systematic way of deriving field equations and boundary conditions for arbitrary mechanical and coupled problems \citep[cf.][]{Maugin:1980lr,Germain:1973mi,Del-Piero:2009pi}. In the present work it is used in the following form: The virtual power of the inertia forces $\scP^{*}_{\ua}$ balances the virtual power $\scP^{*}_{\internal}$ of the internal and $\scP^{*}_{\ext}$ of the external forces acting on any sub-domain $\scS$ of the material body $\scB$ for any admissible virtual velocity field $\vv^{*}$ and virtual rate of order parameter field $\vopv_{\beta}$, i.e.,
\begin{equation}
  \label{eq:PVP}
  \scP^{*}_{\ua} = \scP^{*}_{\internal} + \scP^{*}_{\ext}\,.
\end{equation}
For the sake of simplicity we disregard any higher order inertia terms \cite{Mindlin:1964aa} as well as inertial forces acting on the order parameter, resulting in
\begin{equation}
  \label{eq:power_inertia}
  \scP^{*}_{\ua} = \int_{\scS} \rho \dot{\vv} \cdot \vv^{*} \dV\,.
\end{equation}

The power of internal forces is given by
\begin{equation}
  \label{eq:power_internal}
  \scP^{*}_{\internal} = - \int_{\scS} \left( \tS^{\top}:\tL^{*} + \ttT \tridot \grad{\tL^{*}} - \pi_{\beta}\,\vopv_{\beta} + \vxi_{\beta}\cdot\grad{\vopv_{\beta}} \right) \dV\,,
\end{equation}
with $\tL^{*}:=\grad{\vv^{*}}$. Here $\tS$ and $\ttT$ are the \textsc{Cauchy} and higher order stresses, respectively, while $\pi_{\beta}$ and $\vxi_{\beta}$ are thermodynamic forces that directly correspond to the internal microforce and microstress introduced by \citet{Gurtin:1996vn}. We note that the invariance requirement of $\scP^{*}_{\internal}$ with respect to superimposed rigid body motions is satisfied sufficiently by assuming $\tS = \tS^{\top}$ and $\ttT\cdot\vec{a} = (\ttT\cdot\vec{a})^{\top}$ for arbitrary vectors $\vec{a}$. For the power of external forces we consider the very simple case of no body or contact forces acting on $\tL^{*}$ and $\grad{\vopv_{\beta}}$, and only a contact (micro)force $\zeta_{\beta}$ acting $\vopv_{\beta}$
\begin{equation}
  \label{eq:power_external}
  \scP^{*}_{\ext} = \int_{\scS} \vf \cdot \vv^{*} \rho \dV + \int_{\partial\scS}\left( \vt \cdot \vv^{*} + \zeta_{\beta}\,\vopv_{\beta} \right)\dA\,.
\end{equation}
In order to obtain the consequences of the PVP, the integrals in Eq.~\eqref{eq:power_internal} are transformed using the following identities 
\begin{align}
  &\diver{(\tS\cdot\vv^{*})} = (\diver{\tS})\cdot\vv^{*} + \tS:\tL^{*}\,, \\
  &\diver{(\ttT:\tL^{*})} = (\diver{\ttT}):\tL^{*} + \ttT\tridot\grad{\tL^{*}}\,, \\
  &\diver{\bigl((\diver{\ttT})\cdot\vv^{*}\bigr)} = (\diver{\diver{\ttT}})\cdot\vv^{*} + (\diver{\ttT}):\tL^{*}\,, \\
  &\diver{(\vxi_{\beta}\,\vopv_{\beta})} = (\diver{\vxi_{\beta}})\,\vopv_{\beta} + \vxi_{\beta}\cdot\grad{\vopv_{\beta}}\,,
\end{align}
and the divergence theorem, resulting in
\begin{multline}
  \label{eq:power_internal_1}
  \scP^{*}_{\internal} =  \int_{\scS} \bigl( \diver{\tS} - \diver{\diver{\ttT}}\bigr)\cdot\vv^{*}\dV - \int_{\partial\scS}\vn\cdot\bigl(\tS^{\top} - \diver{\ttT}\bigr)\cdot\vv^{*}\dA - \int_{\partial\scS}\vn\cdot\ttT:\tL^{*}\dA \\
+ \int_{\scS} \bigl( \pi_{\beta} + \diver{\vxi_{\beta}} \bigr)\,\vopv_{\beta} \dV - \int_{\partial\scS}\vn\cdot\vxi_{\beta}\,\vopv_{\beta}\dA\,.
\end{multline}
Introducing the surface gradient operator
\begin{equation}
  \label{eq:surface_gradient}
  \gradS(\cdot) = \grad(\cdot) - \partial_{\vn}(\cdot)\otimes\vn\,, 
\end{equation}
where $\partial_{\vn}$ is the directional derivative in the direction of the outward normal $\vn$, the third integral in expression \eqref{eq:power_internal_1} can be rewritten as
\begin{align}
  \label{eq:third_int}
  \int_{\partial\scS}\vn\cdot\ttT:\tL^{*}\dA   
  &=\int_{\partial\scS}\vn\cdot\ttT:\gradS{\vv^{*}}\dA + \int_{\partial\scS}\vn\cdot\ttT:\partial_{\vn}\vv^{*}\otimes\vn\dA \\
    \nonumber
  &= \int_{\partial\scS}\diverS{\bigl(\vn\cdot\ttT\cdot\vv^{*}\bigr)}\dA
    - \int_{\partial\scS}\diverS{\bigl(\vn\cdot\ttT\bigr)}\cdot\vv\dA + \int_{\partial\scS}\vn\cdot\ttT:\partial_{\vn}\vv^{*}\otimes\vn\dA,.
\end{align}
Finally, applying the surface divergence theorem and, for the sake of simplicity, neglecting any wedge line and corner contributions,  we find 
\begin{equation}
  \label{eq:surf_diver}
  \int_{\partial\scS}\diverS{\bigl(\vn\cdot\ttT\cdot\vv^{*}\bigr)}\dA = \int_{\partial\scS}\bigl( \diverS{\vn} \bigr)\vn\otimes\vn:\ttT\cdot\vv^{*}\dA\,.
\end{equation}
Enforcing Eq.~\eqref{eq:PVP}  we arrive after a number of straight forward algebraic manipulations at the following field equations on $\scB$
\begin{subequations}
\begin{align}
  \label{eq:bal_lin_mom}
  & \rho \dot{\vv} = \diver{\left( \tS - \diver\ttT \right)} + \rho\vf\,,\\
  \label{eq:bal_op}
  & 0 = \diver{\vxi_{\beta}} + \pi_{\beta}\,,
\end{align}
and boundary conditions on $\partial\scB$
\begin{align}
  \label{eq:mech_BC}
  & \vt = \left( \tS - \diver\ttT \right)\cdot\vn - \diverS{\left(\vn\cdot\ttT\right)}\,,\\
  & \zeta_{\beta} = \vxi_{\beta}\cdot\vn\,.
\end{align}
\end{subequations}
We note that, introducing the total stress
\begin{equation}
  \label{eq:total_stress}
  \tSt := \tS - \diver{\ttT}\,,
\end{equation}
the balance of linear momentum \eqref{eq:bal_lin_mom} regains its standard form for simple materials
\begin{align}
  \label{eq:bal_lin_mom_St}
  & \rho \dot{\vv} = \diver{\tSt} + \rho\vf\,,
\end{align}
which is convenient for the numerical implementation.

\section{Constitutive equations}
\label{sec:const-equat}
The following equations are formulated assuming a geometrically linear setting, i.e., the displacement gradient is considered to be small $||\grad{\vu}||\ll 1$. In this case the deformation is characterized by the linear strain tensor $\tE = \half\left( \grad{\vu} + (\grad{\vu})^{\top}\right)$. Its gradient will be denoted by $\ttY := \grad\tE$.

\subsection{Laws of state}
\label{sec:laws-state}
We choose the following ansatz for the specific free energy and thermodynamic forces
\begin{align*}
  & \psi = \psi\left( \tE\,,\,\ttY\,,\, \phi_{\beta}\,,\, \grad\phi_{\beta}\,,\,\theta \right)\,,
  && \tS = \tS\left( \tE\,,\,\ttY\,,\, \phi_{\beta}\,,\, \grad\phi_{\beta}\,,\,\theta \right)\,,
  && \ttT = \ttT\left( \tE\,,\,\ttY\,,\, \phi_{\beta}\,,\, \grad\phi_{\beta}\,,\,\theta \right)\,, \\
  &~
  && \pi_{\beta} = \pi_{\beta}\left( \tE\,,\,\ttY\,,\,\phi_{\beta}\,,\,\grad\phi_{\beta}\,,\,\theta\,,\,\sdot\phi_{\beta} \right)\,,
  && \vxi_{\beta} = \vxi_{\beta}\left( \tE\,,\,\ttY\,,\, \phi_{\beta}\,,\, \grad\phi_{\beta}\,,\,\theta \right)\,.
\end{align*}
The second law of the thermodynamics in the form of the \textsc{Clausius-Duhem} inequality given for the isothermal case by
\begin{equation}
  \label{eq:CD_inequality}
  \left( \tS - \rho\pder{\psi}{\tE} \right):\sdot\tE + \left( \ttT - \rho\pder{\psi}{\ttY} \right)\tridot\sdot\ttY - \left( \pi_{\beta} + \rho\pder{\psi}{\phi} \right)\sdot\phi + \left( \vxi_{\beta} - \rho\pder{\psi}{\grad\phi_{\beta}} \right)\cdot\grad\sdot\phi_{\beta} \geq 0
\end{equation}
can be exploited using the classical \textsc{Coleman-Noll} procedure to arrive at the laws of state
\begin{align}
  \label{eq:laws_of_state}
  & \tS  = \rho\pder{\psi}{\tE}\,, && \ttT = \rho\pder{\psi}{\ttY}\,, && \vxi_{\beta} = \rho\pder{\psi}{\grad\phi_{\beta}}
\end{align}
and the residual dissipation inequality
\begin{align}
  \label{eq:res_dissipation}
  -\pi^{\ud}_{\beta}\,\sdot\phi_{\beta} \geq 0\,, \quad \text{with} \quad \pi^{\ud}_{\beta} := \pi_{\beta} + \rho\pder{\psi}{\phi_{\beta}}\,.
\end{align}

\subsection{Free energy and dissipation potential}
\label{sec:free-energy-dissipation-pot}
As customary in phase field models for solid-solid transformations, the specific free energy can be split into an elastic, a bulk chemical and an interface contribution
\begin{equation}
  \psi = \psi_{\ue}\left( \tE\,,\,\ttY\,,\,\phi_{\beta}\,,\,\theta \right) + \psi_{\ub}\left( \phi_{\beta}\,,\, \theta \right) + \psi_{\ui}\left( \phi_{\beta}\,,\, \grad\phi_{\beta}\,,\, \theta \right)\,,
\end{equation}
as indicated by the subscripts ``e'' (elastic), ``b'' (bulk chemical) and ``i'' (interface). In our formulation, the elastic free energy is of \textsc{Helmholtz}-type, i.e.,
\begin{align}
  \label{eq:helmholtz_type}
  &\rho\psi_{\ue}\left( \tE\,,\,\ttY\,,\,\phi_{\beta}\,,\,\theta \right) =
  \half \tEe : \bbC(\phi_{\beta}) : \tEe
  + \half \bigl( \bbC(\phi_{\beta}):\ttY\cdot\tLcore \bigr) \tridot \ttY\,,\text{ or}\\
  &\rho\psi_{\ue}\left( \tE\,,\,\ttY\,,\,\phi_{\beta}\,,\,\theta \right) =
    \half \bbC^{ijkl}(\phi_{\beta}) E_{ij}^{\text{e}}(\tE\,,\,\phi_{\beta})E_{kl}^{\text{e}}(\tE\,,\,\phi_{\beta})
  + \half \bbC^{ijkl}(\phi_{\beta})\Lambda_{mn}(\phi_{\beta}) \ttY_{ij}^{n}\ttY_{kl}^{m}\,
\end{align}
where $\tEtr$ is the inelastic strain, $\tEe:=\tE - \tEtr$ is the elastic strain, $\bbC(\phi_{\beta})$ the stiffness tensor and $\tLcore$ a gradient length scale tensor \citep[cf.][]{Po:2018}. The specific choice of functional dependence of $\tEtr$, $\psi_{\ub}\left( \phi_{\beta}\,,\, \theta \right)$ and  $\psi_{\ui}\left( \phi_{\beta}\,,\, \grad\phi_{\beta}\,,\, \theta \right)$ on the order parameter $\phi_{\beta}$ is of no relevance at this point; however, we will assume that the interface energy is of the form
\begin{align}
  \label{eq:interface_en_dens}
  &\rho\psi_{\ui}\left( \phi_{\beta}\,,\, \grad\phi_{\beta}\,,\, \theta \right) := \frac{\alpha}{2}\, || \grad\phi_{\beta} ||^{2} + g(\phi_{\beta}\,,\,\theta)\,,
    && \rho\psi_{\ui}\left( \phi_{\beta}\,,\, \grad\phi_{\beta}\,,\, \theta \right) := \frac{\alpha}{2}\, \phi_{\beta}^{,i}\phi_{\beta}^{,i} + g(\phi_{\beta}\,,\,\theta)\,.
\end{align}
Using the laws of state (\ref{eq:laws_of_state}) we immediately find
\begin{subequations}
\begin{align}
  \label{eq:laws_of_state_S}
  & \tS   = \bbC(\phi_{\beta}) : \bigl( \tE - \tEtr \bigr) \,,
  && S^{ij} =  \bbC^{ijkl}(\phi_{\beta})\left(E_{kl} - E_{kl}^{\text{in}}(\phi_{\beta})\right)\,,\\
    \label{eq:laws_of_state_HOS}
  & \ttT = \bbC(\phi_{\beta}):\ttY\cdot\tLcore\,,
  && \ttT^{ij}_{n} = \bbC^{ijkl}(\phi_{\beta})\Lambda_{mn}(\phi_{\beta}) \ttY_{kl}^{m}\,,\\
     \label{eq:laws_of_state_OP}
  & \vxi_{\beta} = \alpha\grad\phi_{\beta}\,,
  && \xi_{\beta}^{i} = \alpha\phi_{\beta}^{,i}\,,
\end{align}
\end{subequations}
and combining the first two equations
\begin{align}
  \label{eq:HOS_in_S}
  &\ttT  = \bbC(\phi_{\beta}):\grad{\left(\bbC^{-1}(\phi_{\beta}):\tS\right)}\cdot\tLcore + \bbC(\phi_{\beta}):\grad{\tEtr}\cdot\tLcore\,, \text{ or}\\
  &\ttT^{ij}_{n} = \bbC^{ijkl}(\phi_{\beta})\Lambda_{mn}(\phi_{\beta})\left(\bbC^{-1}_{klpq}(\phi_{\beta}) S^{pq}\right)^{,m} + \bbC^{ijkl}(\phi_{\beta})\Lambda_{mn}(\phi_{\beta})E_{kl}^{\text{in},m}(\phi_{\beta}) \,.
\end{align}
Equation \eqref{eq:total_stress} can now be used in two ways: In conjunction with the laws of state \eqref{eq:laws_of_state_S} and \eqref{eq:laws_of_state_HOS} it is a constitutive equation for the total stress $\tSt$, which enters the balance of linear momentum \eqref{eq:bal_lin_mom_St}
\begin{align}
  \label{eq:const_St}
  &\tSt\bigl( \tE\,,\, \ttY\,,\, \phi_{\beta} \bigr) = \bbC(\phi_{\beta}):\tEe - \diver{\left[ \bbC(\phi_{\beta}):\ttY\cdot\tLcore\right]}\,, \text{ or}\\
  &S_{\text{t}}^{ij}\bigl( \tE\,,\, \ttY\,,\, \phi_{\beta} \bigr) = \bbC^{ijkl}(\phi_{\beta}):E^{\text{e}}_{kl}\left( \tE\,,\,\phi_{\beta} \right) - \left( \bbC^{ijkl}(\phi_{\beta}):\ttY_{kl}^{m}\Lambda_{mn} \right)^{,n}\,.
\end{align}
When combined with Eq.~\eqref{eq:HOS_in_S}, Eq.~\eqref{eq:total_stress} can  be used  to determine the true stress $\tS$ from the total stress $\tSt$
\begin{align}
  \label{eq:total_stress_S}
  &\tS -\diver{\left[  \bbC(\phi_{\beta}):\grad{\left(\bbC^{-1}(\phi_{\beta}):\tS\right)}\cdot\tLcore \right]} = \tSt + \diver{\bigl(\bbC(\phi_{\beta}):\grad{\tEtr}\cdot\tLcore\bigr)}\,, \text{ or}\\
  &S^{ij} - \left[ \bbC^{ijkl}(\phi_{\beta})\Lambda_{mn}(\phi_{\beta})\left(\bbC^{-1}_{klpq}(\phi_{\beta}) S^{pq}\right)^{,m} \right]^{,n} = S_{\text{t}}^{ij} + \left[ \bbC^{ijkl}(\phi_{\beta})\Lambda_{mn}(\phi_{\beta})E_{kl}^{\text{in},m}(\phi_{\beta}) \right]^{,n}\,.
\end{align}

In order to complete the phase field formulation we require a constitutive equation for $\pi^{\ud}_{\beta}$, which is obtained in the spirit of classical irreversible thermodynamics as
\begin{equation}
  \label{eq:diss_pot}
  \sdot\phi_{\beta} = -\pder{\pdiss}{\pi^{\ud}_{\beta}}
\end{equation}
from a dissipation potential $\pdiss$ that is homogeneous of degree two
\begin{equation}
  \label{eq:diss_pot_expl}
  \pdiss := \half M\left( \pi^{\ud}_{\beta} \right)^{2}\,,
\end{equation}
where $M$ is the so called mobility constant. Combining equations (\ref{eq:bal_op}), (\ref{eq:res_dissipation}), (\ref{eq:laws_of_state_OP}), (\ref{eq:diss_pot}) and (\ref{eq:diss_pot_expl}) we find the classical \textsc{Allen-Cahn} equation
\begin{align}
  \label{eq:AC_equation}
  &M^{-1}\sdot\phi_{\beta} = \alpha\Delta\phi_{\beta} - \rho\pder{\psi}{\phi_{\beta}}\,,
  && M^{-1}\sdot\phi_{\beta} = \alpha\phi_{\beta}^{,ii} - \rho\pder{\psi}{\phi_{\beta}}\,,
\end{align}
or, explicitely writing down the partial derivatives of $\psi$,
\begin{multline}
  \label{eq:AC_equation_explicit}
  M^{-1}\sdot\phi_{\beta} = \alpha\Delta\phi_{\beta} + \tS:\pder{\tEtr}{\phi_{\beta}} - \half\tEe:\pder{\bbC(\phi_{\beta})}{\phi_{\beta}}:\tEe - \half\left( \bbC(\phi_{\beta}):\ttY\cdot\pder{\tLcore}{\phi_{\beta}} \right)\tridot \ttY -\\-  \half\left( \pder{\bbC(\phi_{\beta})}{\phi_{\beta}}:\ttY\cdot\tLcore \right) \tridot \ttY - \rho\pder{\psi_{\ub}(\phi_{\beta}\,,\,\theta)}{\phi_{\beta}} - \pder{g(\phi_{\beta}\,,\,\theta)}{\phi_{\beta}}\,,
\end{multline}
or
\begin{multline*}
M^{-1}\sdot\phi_{\beta} = \alpha\phi_{\beta}^{,ii} + S^{ij}\pder{E^{\text{in}}_{kl}(\phi_{\beta})}{\phi_{\beta}} - \half\pder{\bbC_{ijkl}(\phi_{\beta})}{\phi_{\beta}}E^{\text{e}}_{ij}\left( \tE\,,\,\phi_{\beta} \right)E^{\text{e}}_{kl}\left( \tE\,,\,\phi_{\beta} \right) - \half \pder{\bbC^{ijkl}(\phi_{\beta})}{\phi_{\beta}}\Lambda_{mn}(\phi_{\beta}) \ttY_{ij}^{n}\ttY_{kl}^{m} -\\-  \half \bbC^{ijkl}(\phi_{\beta})\pder{\Lambda_{mn}(\phi_{\beta})}{\phi_{\beta}} \ttY_{ij}^{n}\ttY_{kl}^{m} - \rho\pder{\psi_{\ub}(\phi_{\beta}\,,\,\theta)}{\phi_{\beta}} - \pder{g(\phi_{\beta}\,,\,\theta)}{\phi_{\beta}}\,,
\end{multline*}
Note that all terms that appear in the driving force, and as per \citet{Lazar:2005} the \textsc{Cauchy} stress $\tS$ in particular, are non-singular even in the presence of dislocations. Interestingly, this is not true for an elastic specific free energy that is quadratic in $\ttYe:=\grad\tEe$ rather that $\ttY$ (cf. \ref{sec:why-Ye-is-bad}).

\subsection{Formulation for specific cases}
\label{sec:special-cases}
For phase transformations the crystal lattice on both sides of the interface will, in general, be different leading to different elastic properties and a different shape of the dislocation core. In this case the equations \eqref{eq:bal_lin_mom_St}, \eqref{eq:const_St}, \eqref{eq:total_stress_S} and (\ref{eq:AC_equation_explicit}) retain their full complexity. However, the strength of the general formulation is that it also covers simplified special cases. In the following, we consider scenarios for which these equations can be strongly reduced and which therefore elucidates the structure of the whole formalism.

\subsubsection{Homogeneous bulk material}
\label{sec:homog-bulk-mater}
In the bulk phase the order parameter does not vary in space, i.e., $\grad\phi_{\beta} = \vec{0}$, $\bbC(\phi_{\beta}) = \bbC$, $\tLcore = \tLcorec$, $\tEtr = \vec{0}$. The \textsc{Allen-Cahn} equation is fulfilled automatically  and Eqs.~\eqref{eq:total_stress_S} and \eqref{eq:const_St} recover the form derived by \citet{Po:2018}
\begin{subequations}
 \label{eq:bulk_nl_elast}
\begin{align}
  \label{eq:eqs_bulk}
  & \tS - \diver{\bigl( (\grad{\tS})\cdot\tLcorec} \bigr) = \tSt\,, &\text{with} &&\tSt\bigl( \tE\,,\, \ttY\bigr) = \bbC:\left[ \tE  - \diver{\bigl( \ttY\cdot\tLcorec} \bigr) \right]\,.
\end{align}
For materials with cubic symmetry the gradient length scale tensor $\tLcorec$ is isotropic, i.e., $\tLcorec=\lcorec^{2}\tone$, and the above expressions can be further simplified to the form derived by \citet{Lazar:2005}
\begin{align}
  \label{eq:eqs_bulk_cubic}
  & \tS - \lcorec^{2}\Delta\tS = \tSt\,, &\text{with} &&\tSt\bigl( \tE\,,\, \ttY\bigr) = \bbC:\bigl( \tE  - \lcorec^{2}\diver{\ttY} \bigr)  = \bbC:\bigl( \tE  - \lcorec^{2}\Delta\tE \bigr)\,.
\end{align}
\end{subequations}

\subsubsection{Boundaries between grains without inelastic strain}
\label{sec:grain-boundaries}

The crystal lattices on both sides of a grain boundary differ only by a rotation $\tQ(\phi_{\beta})$. Hence, we assume that the  chemical bulk energy is independent of the order parameter, i.e., $\psi_{\ub}\left( \phi_{\beta}\,,\, \theta \right) = \psi_{\ub}\left( \theta \right)$. Then the elastic stiffness $\bbC(\phi_{\beta})$ and the gradient length scale tensor $\tLcore$ can be expressed as $\bbC(\phi_{\beta}) = \tQ(\phi_{\beta})*\bbC$ and $\tLcore = \tQ(\phi_{\beta})*\tLcorec$, respectively. In the absence of inelastic strain, we have $\tEtr = \vec{0}$. For this case  Eqs.~\eqref{eq:total_stress_S}, \eqref{eq:const_St}  and (\ref{eq:AC_equation_explicit}) take the form
\begin{subequations}
  \label{eq:eqs_grain}
  \begin{align}
    & \tS -  \diver{\left[ \bigl( \tQ(\phi_{\beta})*\bbC \bigr):\grad{\Bigl( \bigl( \tQ(\phi_{\beta})*\bbC^{-1} \bigr):\tS\Bigr)}\cdot\bigl( \tQ(\phi_{\beta})*\tLcorec \bigr) \right]} = \tSt\,,\\
    &\text{with}\nonumber\\
    &\tSt\bigl( \tE\,,\, \ttY\,,\, \phi_{\beta} \bigr) = \bbC(\phi_{\beta}):\tE - \diver{\left[ \bigl( \tQ(\phi_{\beta})*\bbC \bigr):\ttY\cdot\bigl( \tQ(\phi_{\beta})*\tLcorec \bigr) \right]}\,,
  \end{align}
  and
  \begin{multline}
    \label{eq:AC_equation_grain}
    M^{-1}\sdot\phi_{\beta} = \alpha\Delta\phi_{\beta} - \half\tE:\bigl( \pder{\tQ(\phi_{\beta})}{\phi_{\beta}}*\bbC \bigr):\tE - \half\left(  \bigl( \tQ(\phi_{\beta})*\bbC \bigr):\ttY\cdot\bigl( \pder{\tQ}{\phi_{\beta}}*\tLcorec  \bigr)\right)\tridot \ttY -\\-  \half\left( \bigl( \pder{\tQ(\phi_{\beta})}{\phi_{\beta}}*\bbC  \bigr):\ttY\cdot\bigl( \tQ(\phi_{\beta})*\tLcorec \bigr) \right) \tridot \ttY - \pder{g(\phi_{\beta}\,,\,\theta)}{\phi_{\beta}}\,.
  \end{multline}
\end{subequations}
The isotropy of the gradient length scale tensor $\tLcorec$ for cubic crystals implies that $ \tQ(\phi_{\beta})*\tLcorec=\tLcorec=\lcorec^{2}\tone$, which simplifies Eqs.~\eqref{eq:eqs_grain} to the following form
\begin{subequations}
  \begin{align}
    & \tS -  \lcorec^{2}\diver{\left[ \bigl( \tQ(\phi_{\beta})*\bbC \bigr):\grad{\Bigl( \bigl( \tQ(\phi_{\beta})*\bbC^{-1} \bigr):\tS\Bigr)} \right]} = \tSt\,,\\
    &\text{with}\nonumber\\
    &\tSt\bigl( \tE\,,\, \ttY\,,\, \phi_{\beta} \bigr) = \bbC(\phi_{\beta}):\tE - \lcorec^{2}\diver{\left[ \bigl( \tQ(\phi_{\beta})*\bbC \bigr):\ttY \right]}\,,\\
&\text{and}\nonumber\\
    \label{eq:AC_equation_grain_cubic}
     &M^{-1}\sdot\phi_{\beta} = \alpha\Delta\phi_{\beta} - \half\tE:\bigl( \pder{\tQ(\phi_{\beta})}{\phi_{\beta}}*\bbC \bigr):\tE -  \half\lcorec^{2}\left( \bigl( \pder{\tQ(\phi_{\beta})}{\phi_{\beta}}*\bbC  \bigr):\ttY\right) \tridot \ttY - \pder{g(\phi_{\beta}\,,\,\theta)}{\phi_{\beta}}\,.
  \end{align}
\end{subequations}

\subsubsection{Twin boundaries and boundaries between grains with inelastic strain}
\label{sec:twin-boundaries}

Since the twin variants on both sides of the boundary are related by mirror and/or rotational symmetry transformations between the unit cells, we can - as in the case of grain boundaries - assume that the bulk chemical energy  remains unchanged, i.e., $\psi_{\ub}\left( \phi_{\beta}\,,\, \theta \right) = \psi_{\ub}\left( \theta \right)$, and the elastic stiffness $\bbC(\phi_{\beta})$ and the gradient length scale tensor $\tLcore$ can be expressed using an orthogonal tensor $\tQ(\phi_{\beta})$ as  $\bbC(\phi_{\beta}) = \tQ(\phi_{\beta})*\bbC$ and $\tLcore = \tQ(\phi_{\beta})*\tLcorec$, respectively. Under these assumptions we find
\begin{subequations}
  \begin{align}
    \label{eq:total_stress_S_twin}
    & \tS -  \diver{\left[ \bigl( \tQ(\phi_{\beta})*\bbC \bigr):\grad{\Bigl( \bigl( \tQ(\phi_{\beta})*\bbC^{-1} \bigr):\tS\Bigr)}\cdot\bigl( \tQ(\phi_{\beta})*\tLcorec \bigr) \right]} = \tSt + \diver{\left[\bbC(\phi_{\beta}):\grad{\bigl( \tEtr \bigr)\cdot\tLcore}\right]} \,,\\
    &\text{with}\nonumber\\
    \label{eq:const_St_twin}
    &\tSt\bigl( \tE\,,\, \ttY\,,\, \phi_{\beta} \bigr) = \bbC(\phi_{\beta}):\tE_{\ue}(\tE) - \diver{\left[ \bigl( \tQ(\phi_{\beta})*\bbC \bigr):\ttY\cdot\bigl( \tQ(\phi_{\beta})*\tLcorec \bigr) \right]}\,,\\
    &\text{and}\nonumber
  \end{align}
  \vspace{-6ex}
  \begin{multline}
    \label{eq:AC_equation_twin}
    M^{-1}\sdot\phi_{\beta} = \alpha\Delta\phi_{\beta} + \tS:\pder{\tEtr}{\phi_{\beta}} - \half\tEe:\bigl( \pder{\tQ(\phi_{\beta})}{\phi_{\beta}}*\bbC \bigr):\tEe -\\- \half\left(  \bigl( \tQ(\phi_{\beta})*\bbC \bigr):\ttY\cdot\bigl( \pder{\tQ}{\phi_{\beta}}*\tLcorec  \bigr)\right)\tridot \ttY -  \half\left( \bigl( \pder{\tQ(\phi_{\beta})}{\phi_{\beta}}*\bbC  \bigr):\ttY\cdot\bigl( \tQ(\phi_{\beta})*\tLcorec \bigr) \right) \tridot \ttY - \pder{g(\phi_{\beta}\,,\,\theta)}{\phi_{\beta}}\,.
  \end{multline}
\end{subequations}
For cubic lattices these expressions simplify to
\begin{subequations}
  \begin{align}
    & \tS -  \lcorec^{2}\diver{\left[ \bigl( \tQ(\phi_{\beta})*\bbC \bigr):\grad{\Bigl( \bigl( \tQ(\phi_{\beta})*\bbC^{-1} \bigr):\tS\Bigr)} \right]} = \tSt + \lcorec^{2}\diver{\left[\bbC(\phi_{\beta}):\grad{ \tEtr }\right]}\,,\\
    &\text{with}\nonumber\\
    &\tSt\bigl( \tE\,,\, \ttY\,,\, \phi_{\beta} \bigr) = \bbC(\phi_{\beta}):\tEe - \lcorec^{2}\diver{\left[ \bigl( \tQ(\phi_{\beta})*\bbC \bigr):\ttY \right]}\,,\\
    &\text{and}\nonumber
  \end{align}
  \vspace{-6ex}
  \begin{multline}
    \label{eq:AC_equation_twin_cubic}
     M^{-1}\sdot\phi_{\beta} = \alpha\Delta\phi_{\beta} + \tS:\pder{\tEtr}{\phi_{\beta}} - \half\tEe:\bigl( \pder{\tQ(\phi_{\beta})}{\phi_{\beta}}*\bbC \bigr):\tEe -\\-  \half\lcorec^{2}\left( \bigl( \pder{\tQ(\phi_{\beta})}{\phi_{\beta}}*\bbC  \bigr):\ttY\right) \tridot \ttY - \pder{g(\phi_{\beta}\,,\,\theta)}{\phi_{\beta}}\,.
   \end{multline}
 \end{subequations}

\subsubsection{Phase boundaries between cubic phases}
\label{sec:phase-boundaries}
In the case of phase boundaries between different cubic phases the gradient length scale tensor $\tLcorec$ is isotropic on both sides of the interface, even though not necessarily constant across the interface, i.e., $\tLcorec=\lcore^{2}\tone$. This allows us to reduce Eqs.~\eqref{eq:total_stress_S}, \eqref{eq:const_St}  and (\ref{eq:AC_equation_explicit}) to the following form
\begin{subequations}
  \begin{align}
    \label{eq:total_stress_S_cc}
    &\tS -\diver{\left[ \lcore^{2}\, \bbC(\phi_{\beta}):\grad{\bigl(\bbC^{-1}(\phi_{\beta}):\tS\bigr)} \right]} = \tSt + \diver{\bigl(\lcore^{2}\,\bbC(\phi_{\beta}):\grad{\tEtr}\bigr)} \,,\\
    & \text{with}\nonumber\\
    \label{eq:const_St_cc}
    &\tSt\bigl( \tE\,,\, \ttY\,,\, \phi_{\beta} \bigr) = \bbC(\phi_{\beta}):\tEe - \diver{\left( \lcore^{2}\, \bbC(\phi_{\beta}):\ttY\right)}\,,\\
    &\text{and}\nonumber
  \end{align}
   \vspace{-6ex}
  \begin{multline}
    \label{eq:AC_equation_explicit_cubic}
    M^{-1}\sdot\phi_{\beta} = \alpha\Delta\phi_{\beta} + \tS:\pder{\tEtr}{\phi_{\beta}} + \half\tEe:\pder{\bbC(\phi_{\beta})}{\phi_{\beta}}:\tEe - \lcore \pder{\lcore}{\phi_{\beta}} \left( \bbC(\phi_{\beta}):\ttY \right) \tridot \ttY -\\- \frac{\lcore^{2}}{2} \left( \pder{\bbC(\phi_{\beta})}{\phi_{\beta}}:\ttY \right) \tridot \ttY - \rho\pder{\psi_{\ub}(\phi_{\beta}\,,\,\theta)}{\phi_{\beta}} - \pder{g(\phi_{\beta}\,,\,\theta)}{\phi_{\beta}}\,.
  \end{multline}
\end{subequations}

\section{Examples}
\label{sec:examples}
To demonstrate the key properties of the above model numerical simulations using the finite element method are performed using the commercial software ``COMSOL Multiphysics''\footnote{\url{https://www.comsol.com/}}. A uniform mesh with quadratic\footnote{Independent of the chosen shape functions, Comsol does not provide third spatial derivatives of the degrees of freedom. Therefore, in order to obtain the second derivative of strain (third spatial derivative of the displacement), the ``Distributed ODE'' feature is used in order to introduce additional degrees of freedom, corresponding to the second spatial derivatives of the displacement. For this ``Distributed ODE'' linear shape functions are employed.}, quadrilateral elements is used for the domain discretization. The element size is 0.2~nm. Time stepping is performed using the BDF method.
Based on the assumptions of the small perturbation hypothesis\footnote{Both the displacement $\vu$ as well as the displacement gradient are considered to be small, i.e., $|\vu|\ll L$ and $||\grad{\vu}||\ll 1$.} \citep{Maugin:1992uq}, we apply traction boundary conditions to the undeformed geometry whenever required.
We assume elastostatics with an isotropic stiffness tensor $\tet{C}$. Material parameters have been chosen to represent $\alpha$-iron with the values of the elastic constants, $E=200$ GPa, $\nu=0.29$, and the Burger's vector $b=0.285$ nm.
\begin{table}[tbp]
  \begin{tabular}{lll}
    \hline
    parameter name & symbol & value\\
    \hline
    Young's modulus & $E$ & 200 GPa \\
    Poisson's ratio & $\nu$ & 0.29 \\
    Burger's vector & $b$ & 0.285 nm\\
    \hline
    coefficient & $a$ & 2.98 \\
    coefficient & $A$ & $1.155\times 10^{8} \text{ J/m}^{3}$\\
    coefficient & $B$ & $-3.43\times 10^{7} \text{ J/m}^{3}$\\
    coefficient & $C$ & $-2.78\times 10^{8} \text{ J/m}^{3}$\\
    mobility & $M$ & $2 \text{ m}^{3}/\text{Js}$\\
    gradient coefficient & $\alpha$ & $5\times 10^{-11} \text{ N}$
  \end{tabular}
  \caption{Model parameters used for the numerical example in Sec.~\ref{sec:MB_interaction}.}
  \label{tab:pars}
\end{table}

\subsection{Regularization in the dislocation core}
\label{sec:MB_reg_core}

\begin{figure}[h]
  \centering
  \subfloat[Shear stress component $S_{12}$ in the glide plane.]{
    \begin{overpic}[percent,scale=0.95]{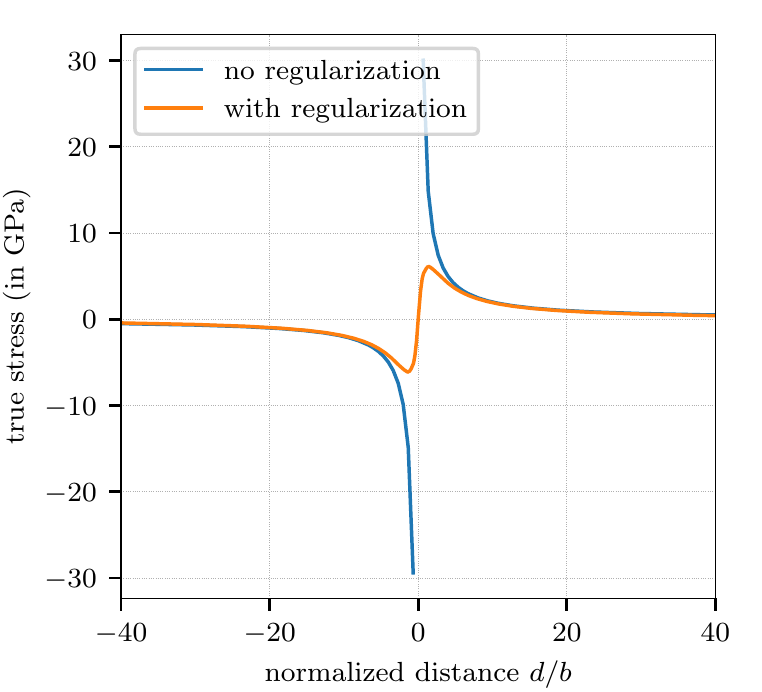}
      \put(62,15){\includegraphics[scale=0.6]{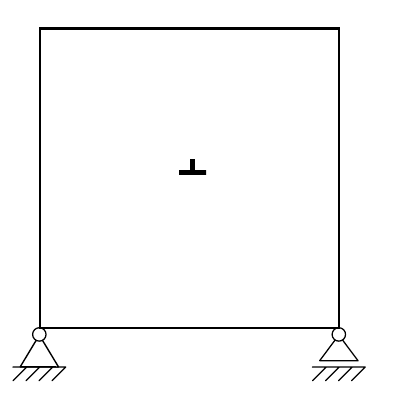}}
    \end{overpic}
    \label{fig:disl}
  }
  \hspace{0.1cm}
  \subfloat[Density plot the of stress component $S_{12}$.]{
    \includegraphics[width=0.45\textwidth]{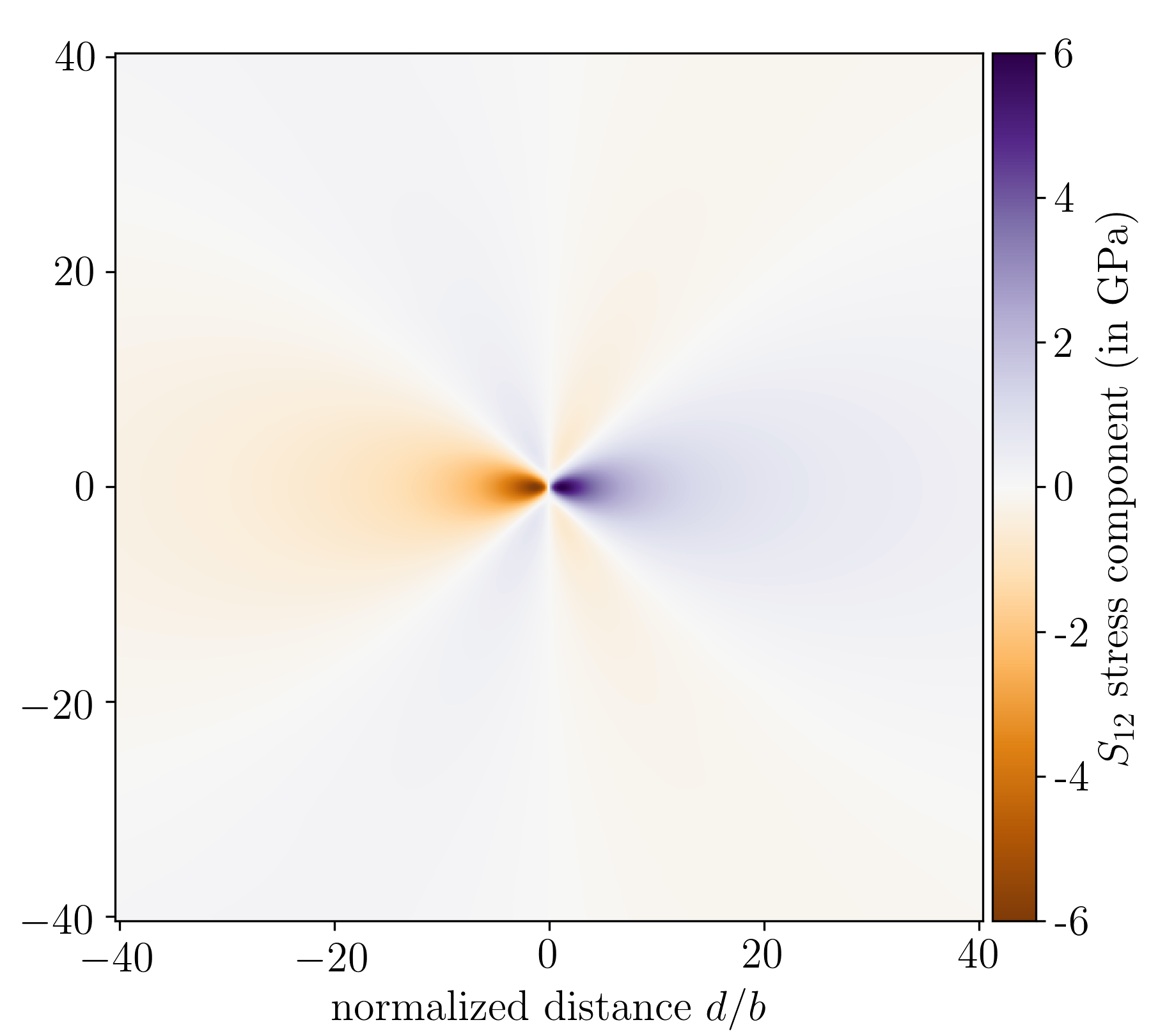}
  }
  \caption{Shear stress component $S_{12}$ for a single edge dislocation. The inset in (a)  shows the simulation setup.}
  \label{fig:MB_single_edge}
\end{figure}

As shown in Sec.~\ref{sec:homog-bulk-mater}, the present model reduces to the set of equations proposed by \cite{Po:2018} in the homogeneous bulk phase. Here, we apply this formulation to a single edge dislocation in an infinite elastic medium: Fig.~\ref{fig:MB_single_edge} shows the shear stress component $S_{12}$ in the plane perpendicular of this dislocation with and without regularization ($\lcorec = 2$ \AA). In the ``classical'' case without regularization, the stress in the dislocation core is singular, whereas it is well defined and finite for the regularized solution, in analogy to what one would expect from a real atomistic configuration.

\subsection{Effect of the regularization on the interaction of dislocations with a moving interface}
\label{sec:MB_interaction}

This examples demonstrates the interaction of dislocations with a moving interface between phase variant 1 (indicated by the superscript ``V1'') and variant 2 (indicated by the superscript ``V2''). The phase mesostructure is described by one single order parameter $\phi$. The only difference between the two variants is with respect to the eigenstrain induced by the phase transformation.  This inelastic strain is given as a function of the order parameter by
\begin{equation}
  \label{eq:in-ex}
  \tEtr(\phi) =
  \begin{cases}
    \tEtr^{\text{V1}}\,\varphi(\phi) & \textrm{if}\;\;\phi \geqslant 0\,,\\
    \tEtr^{\text{V2}}\,\varphi(\phi) & \textrm{if}\;\;\phi < 0\\
  \end{cases}\,,
\end{equation}
where $\tEtr^{\text{V1}}$ and $\tEtr^{\text{V2}}$ are the eigenstrains of the phases V1 and V2, respectively,
\begin{align}
  \label{eq:eig-ex}
  & \tEtr^{\text{V1}} =
    \left(
    \begin{matrix}
      0 & 0.076 \\
      0.076 & 0
    \end{matrix}
            \right)\,,
        && \tEtr^{\text{V2}} =
           \left(
           \begin{matrix}
             0 & -0.076 \\
             -0.076 & 0
           \end{matrix}
                   \right)\,.
\end{align}
$\varphi(\phi)$ is a polynomial chosen in accordance with \cite{Levitas:2002xy}
\begin{equation}
  \label{eq:phi6}
  \varphi(\phi) = \frac{a}{2}\phi^{2} + (3-a)\phi^4 + \half(a-4)\phi^6\,.
\end{equation}
The symmetric bulk chemical free energy takes the following form
\begin{equation}
  \label{eq:bulk-chem-ex}
  \rho\psi_{\ub}(\phi) = A\phi^{6} + B\phi^{4} + C\phi^{2}\,,
\end{equation}
and the interface energy density is assumed as
\begin{equation}
  \label{eq:interface_en_dens}
  \rho\psi_{\ui}\left( \grad\phi \right) = \frac{\alpha}{2}\, || \grad\phi ||^{2}.
\end{equation}
For this specific case the resulting set of partial differential equations \eqref{eq:bal_lin_mom_St} and  \eqref{eq:total_stress_S_cc} can be further simplified to
\begin{subequations}
  \begin{align}
    \label{eq:total_stress_S_ex}
    & \diver{\tSt} = \vec{0}\,,\\
    &\tS -\lcorec^{2}\Delta\tS = \tSt + \lcorec^{2}\,\bbC:\Delta\tEtr \,,\\
    & \text{with}\nonumber\\
    \label{eq:const_St_ex}
    &\tSt\bigl( \tE\,,\, \ttY\,,\, \phi \bigr) = \bbC:\left(  \tE-  \tEtr  \right) - \lcorec^{2}\, \bbC:\diver{ \ttY}\,,\\
    &\text{and}\nonumber\\
    \label{eq:AC_equation_explicit_ex}
    & M^{-1}\sdot\phi = \alpha\Delta\phi + \tS:\pder{\tEtr}{\phi}  - \rho\pder{\psi_{\ub}(\phi)}{\phi} \,.
  \end{align}
\end{subequations}
These equations are solved for the displacement field $\vec{u}$, the order parameter $\phi$ and the true stress $\tS$. All parameters and coefficients occurring  in the above equations are summarized in Tab.~\ref{tab:pars}. The resulting interface energy, computed for a stationary flat interface, is $\gamma = 0.22 \text{ J/m}^2$. The timescale in the simulation is controlled by the mobility constant $M$. Since the simulation time can be arbitrarily re-scaled using the mobility, in our simulations we treat it as dimensionless pseudo time.

\begin{figure}[tbp]
  \centering
  \subfloat[Schematic representation of interface and dislocation arrangement. The system is assumed to be periodic in vertical direction. Only the domain indicated by the dashed box is used in the simulations. The false-color plot indicated domains with positive (red) and negative (blue) in-plane shear stress $S_{12}$.]{
    \def\svgwidth{0.4\textwidth}
    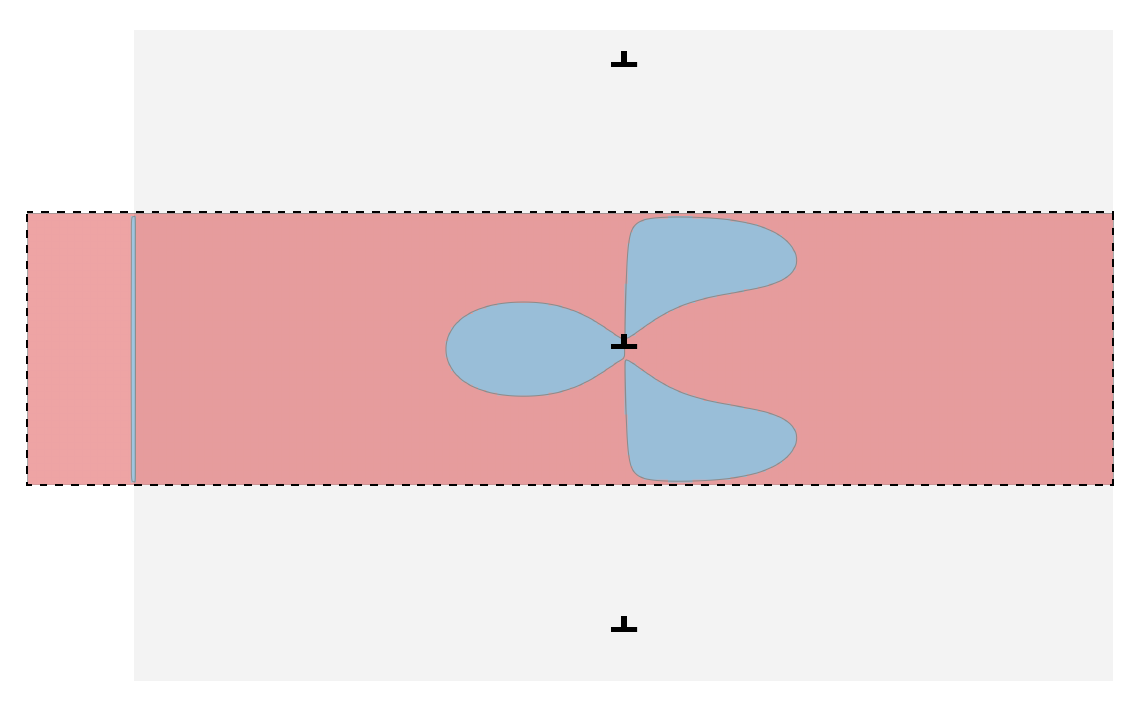
    \label{fig:mm_schem}
  }
  \hspace{1cm}
  \subfloat[In-plane shear stress $S_{12}$ due to the dislocation for the two different regularization lengths $\lcorec$ used in this example.]{
    \includegraphics[width=0.335\textwidth]{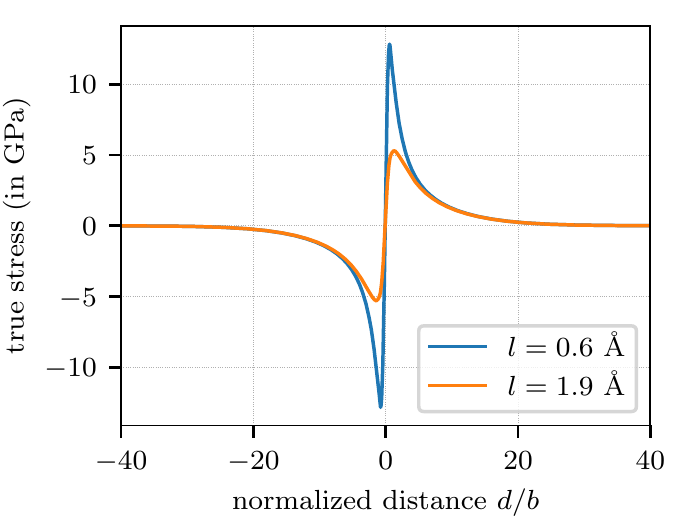}
    \label{fig:mm_stress}
  }
  \caption{Model problem with initially flat phase boundary driven by pure shear loading towards a periodic arrangement of dislocations. }
  \label{fig:mm}
\end{figure}

\begin{figure}[tbp]
  \centering
  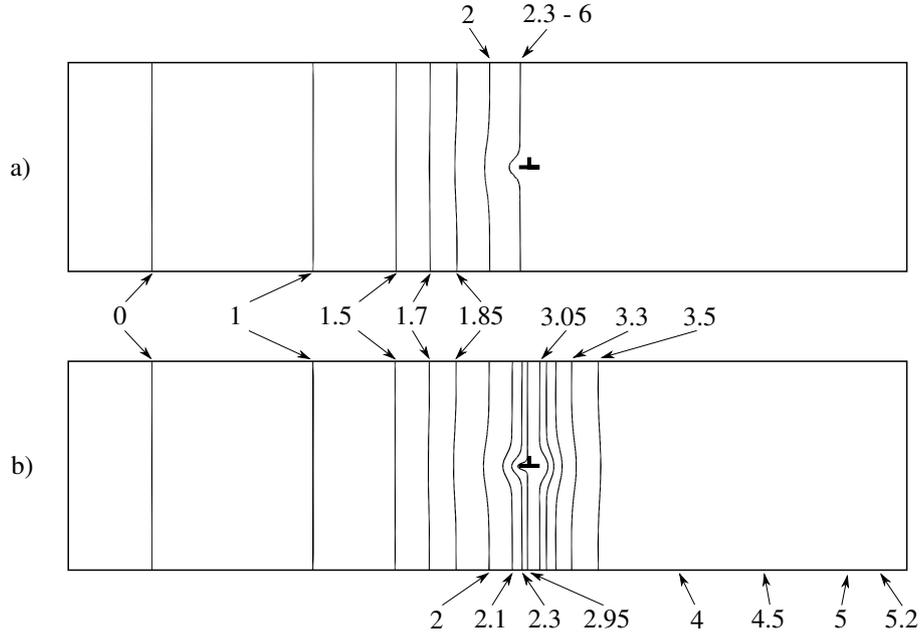
  \caption{Propagation of the interface. The labeled lined correspond to the center of the interface at the pseudo-times (in $\upmu$s) denoted by the corresponding labels. a) $\lcorec = 0.6$ \AA: The interface is arrested at the dislocation array.  b) $\lcorec = 1.9$ \AA: The interface sweeps over the dislocation array.}
  \label{fig:mm_inter}
\end{figure}

\begin{figure}[tbp]
  \centering
    \subfloat[Evolution of the V1 phase content for different regularization lengths $\lcorec$. For $\lcorec = 0.6$ \AA{} the interface is arrested, whereas for $\lcorec = 1.9$ \AA{} it moves past the dislocation array.]{
    \includegraphics[width=0.32\textwidth]{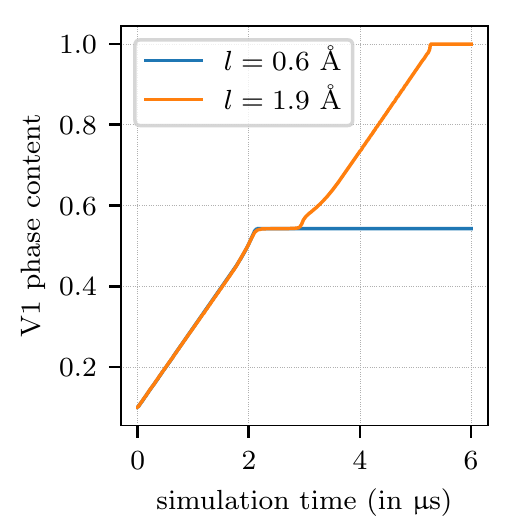}
    \label{fig:mm_ph_overall}
  }
  \hspace{1cm}
  \subfloat[Rate of the evolution of the V1 phase content for different regularization lengths $\lcorec$. The large rate at time 5.2 is an artifact of approaching the boundary of the simulation domain.]{
    \includegraphics[width=0.32\textwidth]{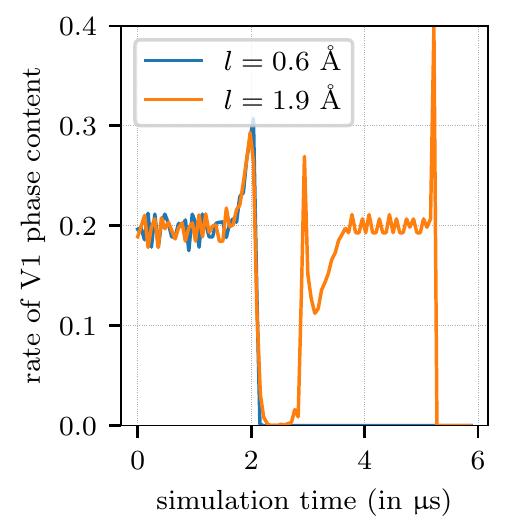}
    \label{fig:mm_ph_overall_vel}
  }\\
  \subfloat[Phase-boundary positions at equidistant time intervals for regularization lengths $\lcorec = 0.6$ \AA{} (left) and $\lcorec = 1.9$ \AA{} (right). An increasing distance between the contours indicates an acceleration of the interface, while a decreasing distance indicates deceleration. The interface positions at times 1, 2, 3, 4, 5, 6 $\upmu$s are shown in red (from left to right). ]{
      \includegraphics[width=0.32\textwidth]{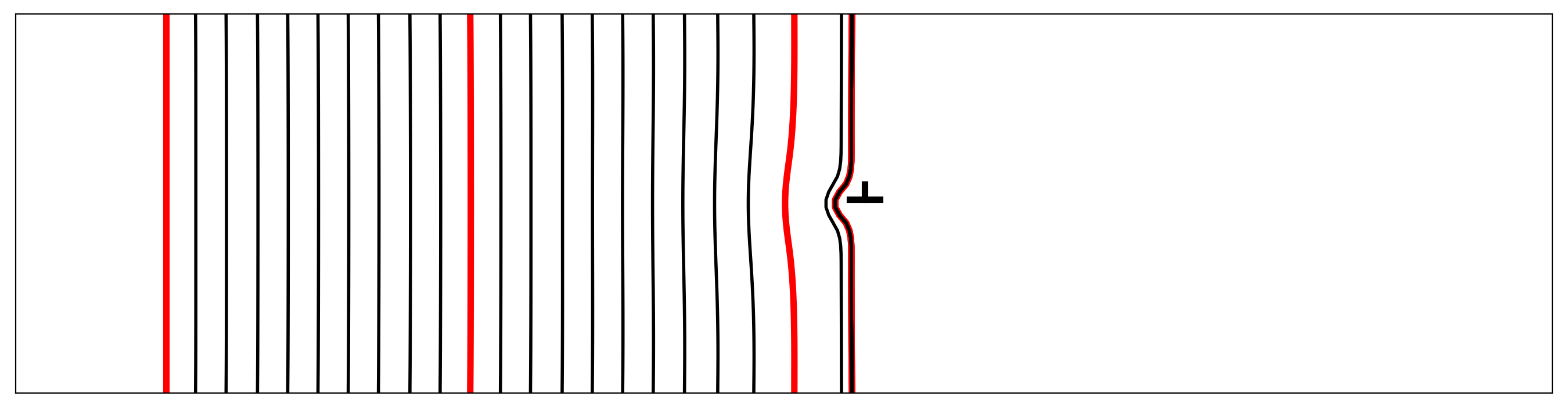}\hspace{1cm}
      \includegraphics[width=0.32\textwidth]{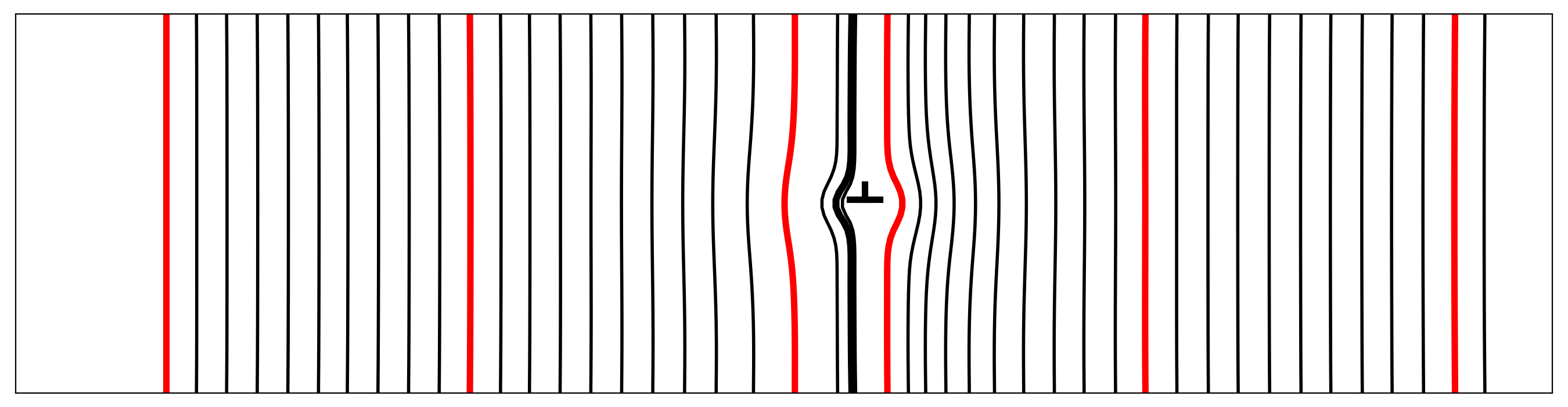}
    \label{fig:mm_ph_contours}
  }\\
  \subfloat[Evolution of the V1 phase content for $\lcorec = 0.6$ \AA{}. A comparison between the overall phase content, the phase content along the centerline of the simulation box and the top of the simulation box.]{
    \includegraphics[width=0.32\textwidth]{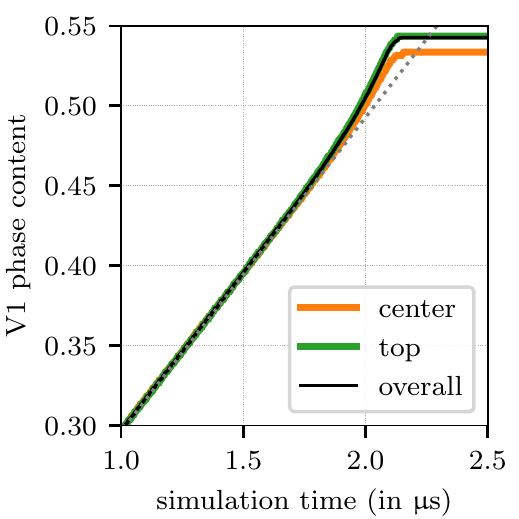}
    \label{fig:mm_ph_6e11}
  }
  \hspace{1cm}
  \subfloat[Evolution of the V1 phase content for $\lcorec = 1.9$ \AA{}. A comparison between the overall phase content, the phase content along the centerline of the simulation box and the top of the simulation box.]{
    \includegraphics[width=0.32\textwidth]{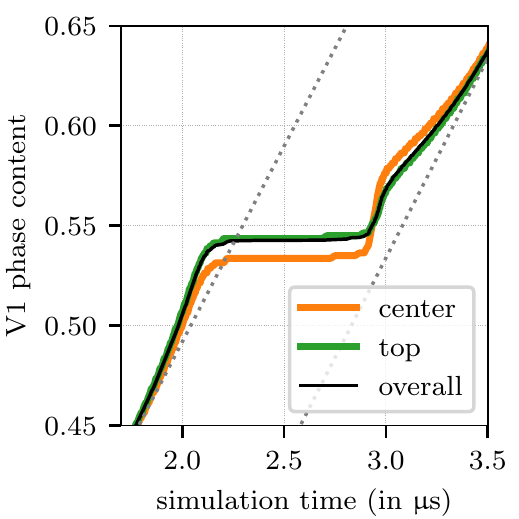}
    \label{fig:mm_ph_19e11}
  }
  \caption{Evolution of the phase content of variant 1 over pseudo-time.}
  \label{fig:mm_ph}
\end{figure}

The following scenario considers an initially flat interface between variants V1 and V2, and a periodic, immobile dislocation structure with a dislocation spacing of 10~nm within variant 2. For a pictorial representation, see Fig.~\ref{fig:mm_schem}. The structure is assumed to be infinite in vertical direction, allowing us to reduce the simulation domain to the dashed 40~nm wide and 10~nm high box in Fig.~\ref{fig:mm_schem} with periodic boundary conditions in vertical direction. The domain is loaded under pure shear conditions with an in-plain shear stress of 85~MPa, under which V1 is energetically more favorable, i.e., the interface will move to the right.

Simulations are carried out using different regularization lengths $\lcorec$ ($\lcorec=0.6$ \AA{} and $\lcorec=1.9$ \AA{}) resulting in different peak stresses in the dislocation core (see Fig.~\ref{fig:mm_stress}).
Fig.~\ref{fig:mm_inter} shows the positions of the V1-V2 interface for different points in simulation time. As the interface approaches the dislocations it bows out due to the interaction with the stress field of the dislocation core. The smaller regularization length results in a larger stress magnitude in the dislocation core region, leading to an arrest of the interface (see Fig.~\ref{fig:mm_inter}a). For the larger regularization length the stress in the vicinity of the dislocation is low enough in order to allow the interface to pass over the dislocation as shown in Fig.~\ref{fig:mm_inter}b. 

To analyze the temporal evolution in more detail Fig.~\ref{fig:mm_ph} visualizes a number of different aspects of the investigated system. The overall V1 phase content, i.e., the area containing phase variant V1 divided by the whole area, as a function of (pseudo) time is shown in Fig.~\ref{fig:mm_ph_overall} for the two different regularization lengths. There, the most obvious characteristic is the arrest of the interface for both values of $\lcorec$ happening simultaneously shortly after $t=2$\,\textmu s. While the system with the smaller regularization length has already reached a stationary state, the other system shows that the interface ``detaches'' from the dislocation and swipes the same area per time as before, which shows in the same inclination of the respective line in Fig.~\ref{fig:mm_ph_overall}. 

How does the  \emph{rate} of the V1 phase content evolution change shortly before and after the arresting of the interface took place? In Fig.~\ref{fig:mm_ph_overall} the peaks at $t\approx 2$ and $\approx 3$\,\textmu s indicate that the phase boundary is accelerating towards the dislocation until its velocity is significantly reduced in the vicinity of the dislocation. The second peak shows that the interface effectively accelerates again after passing the dislocation. This stage is followed by another dip (in between $\approx 3$ and $\approx 3.5$\,\textmu s) where the interface motion right of the dislocation is again decelerated. This behaviors is also visualized in Fig.~\ref{fig:mm_ph_contours}, which shows the interface position at equidistant points in time. 

Two different phenomenons operate here, which can be understood from the change of sign of the shear stress field of an edge dislocation as shown in Fig.~\ref{fig:mm_schem}. Recall that the inelastic strain of the interface is governed only by the shear components of the strain tensor. Once the phase boundary is getting close enough to interact with the dislocation,  the upper and lower sections of phase V2 are in the regions 3/6 of the dislocation (compare Fig.~\ref{fig:mm_schem}). The driving force is effectively directed in positive x-direction and causes the acceleration. The central regions of the phase boundary, is located in region 4 of the dislocation stress field and therefore experiences a net driving force that is directed in \emph{opposite} direction. This interplay between the directions of the two driving forces is also responsible for the curvature of the interface. 
Once the interface has passed the dislocation, the central regions of the interface experiences a very large driving force in positive direction (region 1 of the dislocation). The top and bottom sections of the interface, however, are located in regions with negative driving force 2/6. 
When the interface moves further towards the right, the magnitude of the driving force from the dislocation acting in section 1 decreases as $1/r$ and this region of the interface decelerates. At the same time, the driving force acting on the top and bottom of the interface increases only slightly,  explaining the second dip in Fig~\ref{fig:mm_ph_overall_vel} around $t=3...3.75$\textmu s. At a sufficient distance from the dislocation the top and bottom parts of the interface accelerate, which leads to a decrease in curvature of the interface. 

 This is further shown in Figs.~\ref{fig:mm_ph_6e11} and \ref{fig:mm_ph_19e11}, which relate the motion of the curved interface to the motion of a flat interface for the case without a dislocation structure (indicated by the dotted line). 

\section{Summary}
\label{sec:summary}

In this paper we developed a framework for coupling a phase-field description of planar defects such as phase or twin-boundaries with a discrete representation of dislocations within (anisotropic) first-strain-gradient elasticity. Its main features and advantages in contrast to phase-field within classical elasticity are:
\begin{itemize}
\item Non-singular stresses at the dislocation core that can be easily calibrated to match molecular statics predictions using the approach of \cite{Admal:2017}
\item Non-singular driving forces for the evolution of the phase-field evolution in the presence of dislocation. This ensures a mesh-independent numerical solution and is a necessary condition for modeling the \emph{interaction} of dislocations with interfaces such as phase-, grain- or twin-boundaries. 
\end{itemize}

We have shown that in order to ensure regularized driving forces in the dislocation core, a \textsc{Helmholtz}-type elastic free energy that is quadratic in the gradient of the total rather than the elastic strain must be used.

We implemented the proposed framework in the Comsol Multiphics Modeling Software and demonstrated its feasibility and basic properties based on a number of examples.
Coupled to a dislocation-dynamics code, we expect this phase-field framework to be a valuable tool for understanding microstructure-evolution on a small scale.\\

\textbf{Acknowledgements}\\
The authors gratefully acknowledge the Deutsche Forschungsgemeinschaft (DFG) for supporting this work carried out within the framework of Collaborative Research Center SFB 799. SS acknowledges financial support from the European Research Council through the ERC Grant Agreement No. 759419 MuDiLingo (“A Multiscale Dislocation Language for Data-Driven Materials Science”).

\appendix

\section{Energy that is quadratic in $\ttYe$}
\label{sec:why-Ye-is-bad}
\begin{figure}[tbp]
  \includegraphics[]{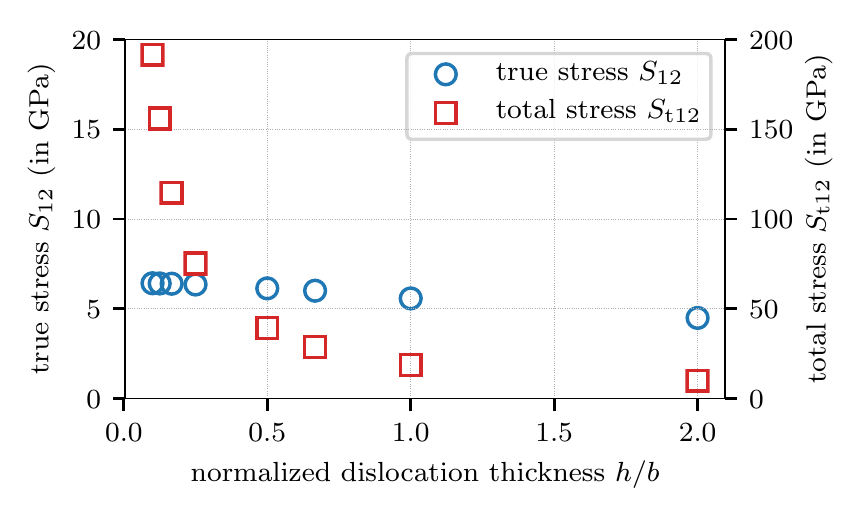}
  \caption{The maximum shear stress in the dislocation core as a function of the dislocation ``thickness''.}
  \label{fig:disl_reg}
\end{figure}

Starting with a setup identical to Sec.~\ref{sec:free-energy-dissipation-pot} but for the \textsc{Helmholtz}-type elastic free energy
\begin{equation}
  \label{eq:helmholtz_type_Ye}
  \rho\psi_{\ue}\left( \tE\,,\,\ttY\,,\,\phi_{\beta}\,,\,\theta \right) =
  \half \tEe : \bbC(\phi_{\beta}) : \tEe
  + \half \left( \bbC(\phi_{\beta}):\ttYe\cdot\tLcore \right) \tridot \ttYe\,,
\end{equation}
we find
\begin{align}
  \label{eq:laws_of_state_S_Ye}
  & \tS  = \bbC(\phi_{\beta}) : \tEe = \bbC(\phi_{\beta}) : \left( \tE - \tEtr \right) \,,\\
    \label{eq:laws_of_state_HOS_Ye}
  &  \ttT = \bbC(\phi_{\beta}):\ttYe\cdot\tLcore\,,\\
     \label{eq:laws_of_state_OP_Ye}
  & \vxi_{\beta} = \alpha\grad\phi_{\beta} - \ttT:\pder{\tEtr}{\phi_{\beta}}\,,
\end{align}
and once again combining the first two equations
\begin{align}
  \ttT = \bbC(\phi_{\beta}):\grad{\left(\bbC^{-1}(\phi_{\beta}):\tS\right)}\cdot\tLcore\,.
\end{align}
From Eq.~\eqref{eq:total_stress} we find the  constitutive equation for the total stress $\tSt$
\begin{equation}
  \label{eq:const_St_Ye}
  \tSt\bigl( \tE\,,\, \ttY\,,\, \phi_{\beta} \bigr) = \bbC(\phi_{\beta}):\tEe - \diver{\left[ \bbC(\phi_{\beta}):\bigl( \ttY - \grad{\tEtr} \bigr)\cdot\tLcore\right]}\,.
\end{equation}
the equation to determine true stress $\tS$ from the total stress $\tSt$
\begin{equation}
  \label{eq:total_stress_S_Ye}
  \tS -\diver{\left[  \bbC(\phi_{\beta}):\grad{\left(\bbC^{-1}(\phi_{\beta}):\tS\right)}\cdot\tLcore \right]} = \tSt\,.
\end{equation}
The evolution equation for the order parameter obtained using the same procedure as in Sec.~\ref{sec:free-energy-dissipation-pot} is
\begin{equation}
  \label{eq:AC_equation_Ye}
  M\sdot\phi_{\beta} = \alpha\Delta\phi_{\beta} + \diver{\left[\ttT:\pder{\tEtr}{\phi_{\beta}}\right]} -  \rho\pder{\psi}{\phi_{\beta}}\,.
\end{equation}
The divergence on the right hand side of \eqref{eq:AC_equation_Ye} is easily evaluated:
\begin{align*}
  \diver{\left[\ttT:\pder{\tEtr}{\phi_{\beta}}\right]}
  &= -\diver{\left[\ttT:\pder{\tEe}{\phi_{\beta}}\right]}\\
  &= -\diver{\ttT}:\pder{\tEe}{\phi_{\beta}} - \ttT\tridot \pder{\ttYe}{\phi_{\beta}}\\
  &= \tSt:\pder{\tEtr}{\phi_{\beta}} -\tS:\pder{\tEe}{\phi_{\beta}} - \ttT\tridot \pder{\ttYe}{\phi_{\beta}}\\
  &= \tSt:\pder{\tEtr}{\phi_{\beta}} + \rho\pder{\psi}{\phi_{\beta}} - \half\tEe:\pder{\bbC(\phi_{\beta})}{\phi_{\beta}}:\tEe - \\
  & \hspace{2.5cm} \half\left( \bbC(\phi_{\beta}):\ttYe\cdot\pder{\tLcore}{\phi_{\beta}} \right)\tridot \ttYe -\\
  & \hspace{3.5cm} \half\left( \pder{\bbC(\phi_{\beta})}{\phi_{\beta}}:\ttYe\cdot\tLcore \right) \tridot \ttYe
\end{align*}
Finally, we find the expression
\begin{multline}
  \label{eq:AC_equation}
  M\sdot\phi_{\beta} = \alpha\Delta\phi_{\beta} + \tSt:\pder{\tEtr}{\phi_{\beta}} - \half\tEe:\pder{\bbC(\phi_{\beta})}{\phi_{\beta}}:\tEe -\\ \quad \half\left( \bbC(\phi_{\beta}):\ttYe\cdot\pder{\tLcore}{\phi_{\beta}} \right)\tridot \ttYe - \half\left( \pder{\bbC(\phi_{\beta})}{\phi_{\beta}}:\ttYe\cdot\tLcore \right) \tridot \ttYe -\\ \rho\pder{\psi_{\ub}(\phi_{\beta}\,,\,\theta)}{\phi_{\beta}} - \pder{g(\phi_{\beta}\,,\,\theta)}{\phi_{\beta}}\,.
\end{multline}
where the total stress $\tSt$ appears in the driving force. In general, this stress cannot be assumed to be bounded in the dislocation-core. This is illustrated in Fig.~\ref{fig:disl_reg} that shows the maximum shear stress in the dislocation core for different ``thicknesses'' of the dislocation, i.e., different discretizations. While the true stress $S_{12}$ does not change noticeably once the discretization is sufficiently fine, the total stress $S_{\text{t}12}$ keeps increasing with decreasing thickness of the dislocation.

\bibliographystyle{elsarticle-harv}
\bibliography{\string~/Documents/Literature/MyLibrary}

\end{document}

%% file: mypreamble.tex
\usepackage[utf8x]{inputenc}
\usepackage[T1]{fontenc}
\usepackage{amssymb,upgreek}
\usepackage[fleqn]{amsmath}
\usepackage{stmaryrd,mathtools,mathrsfs}
\usepackage{dutchcal}
\usepackage{xspace}
\usepackage{accents}
\usepackage{letltxmacro}
\usepackage{todonotes}
\usepackage{multirow}
\usepackage{pifont}
\usepackage[percent]{overpic}
\usepackage{mathcomp,xfrac}
\usepackage[normalem]{ulem}
\usepackage{dblfloatfix,pbox}
\usepackage{floatrow}
\newfloatcommand{capbtabbox}{table}[][\FBwidth]
\usepackage{blindtext}
\usepackage[pdftex,colorlinks,breaklinks]{hyperref}

\usepackage{color,xcolor}
\definecolor{shadecolor}{RGB}{255,255,255}

\graphicspath{{./figs/}}

%% file: symbolcommands.tex
\newcommand{\half}{\frac{1}{2}}

\renewcommand{\vec}[1]{\boldsymbol{#1}}
\newcommand{\tria}[1]{\boldsymbol{\mathcal{#1}}}

\newcommand{\tet}[1]{ \boldsymbol{\mathbb{#1}}}

\newcommand{\ud}{\text{d}}

\newcommand{\pder}[2]{\partial_{#2} #1}

\DeclareMathOperator*{\grad}{grad}
\DeclareMathOperator*{\gradS}{grad_{S}}
\DeclareMathOperator*{\diver}{div}
\DeclareMathOperator*{\diverS}{div_{S}}

\newcommand{\sdot}[1]{\skew{5}\dot{ #1 }}
\newcommand{\tr}[1]{\text{tr} #1}

\renewcommand{\geq}{\geqslant}

\newcommand{\tridot}{\,\smash\vdots\,}

\newcommand{\ub}{\text{b}}

\newcommand{\ui}{\text{i}}
\newcommand{\ue}{\text{e}}
\newcommand{\ua}{\text{a}}

\newcommand{\utr}{\text{in}}
\newcommand{\ut}{\text{t}}

\newcommand{\scB}{\mathscr{B}}
\newcommand{\scS}{\mathscr{S}}
\newcommand{\scP}{\mathscr{P}}

\newcommand{\vt}{\vec{t}}
\newcommand{\vf}{\vec{f}}
\newcommand{\vn}{\vec{n}}
\newcommand{\tone}{\vec{I}}
\newcommand{\vu}{\vec{u}}
\newcommand{\vv}{\vec{v}}

\newcommand{\vxi}{\vec{\xi}}
\newcommand{\tL}{\vec{L}}

\newcommand{\tB}{\vec{B}}
\newcommand{\tE}{\vec{E}}
\newcommand{\tEtr}{\vec{E}^{\utr}(\phi_{\beta})}

\newcommand{\tEe}{\vec{E}^{\ue}\left(\tE\,,\,\phi_{\beta}\right)}
\newcommand{\tS}{\vec{S}}
\newcommand{\tSt}{\vec{S}_{\ut}}
\newcommand{\ttT}{\tria{T}}
\newcommand{\ttY}{\tria{Y}}
\newcommand{\ttYe}{\tria{Y}^{\ue}\left(\ttY\,,\,\phi_{\beta}\right)}

\newcommand{\tQ}{\vec{Q}}
\newcommand{\tA}{\vec{A}}

\newcommand{\lcore}{\mathcal{l}(\phi_{\beta})}
\newcommand{\lcorec}{\mathcal{l}}
\newcommand{\tLcore}{\vec{\Lambda{}}(\phi_{\beta})}
\newcommand{\tLcorec}{\vec{\Lambda{}}}

\newcommand{\pdiss}{\Omega\left(\pi^{\ud}_{\beta}\right)}

\newcommand{\bbC}{\mathbb{C}}

\newcommand{\vopv}{\sdot\phi^*}

\newcommand{\dV}{\,\ud V}
\newcommand{\dA}{\,\ud a}

\newcommand{\ext}{\text{ext}}
\newcommand{\internal}{\text{int}}

%% file: per_disl_schem.pdf_tex
%% Accompanies image file 'per_disl_schem.pdf' (pdf, eps, ps)
\begingroup%
  \makeatletter%
  \providecommand\color[2][]{%
    \errmessage{(Inkscape) Color is used for the text in Inkscape, but the package 'color.sty' is not loaded}%
    \renewcommand\color[2][]{}%
  }%
  \providecommand\transparent[1]{%
    \errmessage{(Inkscape) Transparency is used (non-zero) for the text in Inkscape, but the package 'transparent.sty' is not loaded}%
    \renewcommand\transparent[1]{}%
  }%
  \providecommand\rotatebox[2]{#2}%
  \newcommand*\fsize{\dimexpr\f@size pt\relax}%
  \newcommand*\lineheight[1]{\fontsize{\fsize}{#1\fsize}\selectfont}%
  \ifx\svgwidth\undefined%
    \setlength{\unitlength}{325.98425197bp}%
    \ifx\svgscale\undefined%
      \relax%
    \else%
      \setlength{\unitlength}{\unitlength * \real{\svgscale}}%
    \fi%
  \else%
    \setlength{\unitlength}{\svgwidth}%
  \fi%
  \global\let\svgwidth\undefined%
  \global\let\svgscale\undefined%
  \makeatother%
  \begin{picture}(1,0.62608696)%
    \lineheight{1}%
    \setlength\tabcolsep{0pt}%
    \put(0,0){\includegraphics[width=\unitlength,page=1]{per_disl_schem.pdf}}%
    \put(0.14941818,0.30079054){\color[rgb]{0,0,0}\makebox(0,0)[lt]{\lineheight{1.25}\smash{\begin{tabular}[t]{l}variant 2\end{tabular}}}}%
    \put(0.08079267,0.23543675){\color[rgb]{0,0,0}\rotatebox{90}{\makebox(0,0)[lt]{\lineheight{1.25}\smash{\begin{tabular}[t]{l}variant 1\end{tabular}}}}}%
    \put(0,0){\includegraphics[width=\unitlength,page=2]{per_disl_schem.pdf}}%
    \put(0.62402901,0.3064469){\color[rgb]{0,0,0}\makebox(0,0)[lt]{\lineheight{1.25}\smash{\begin{tabular}[t]{l}\textit{1}\end{tabular}}}}%
    \put(0.62132947,0.38738775){\color[rgb]{0,0,0}\makebox(0,0)[lt]{\lineheight{1.25}\smash{\begin{tabular}[t]{l}\textit{2}\end{tabular}}}}%
    \put(0.45997238,0.38738775){\color[rgb]{0,0,0}\makebox(0,0)[lt]{\lineheight{1.25}\smash{\begin{tabular}[t]{l}\textit{3}\end{tabular}}}}%
    \put(0.45855514,0.30636252){\color[rgb]{0,0,0}\makebox(0,0)[lt]{\lineheight{1.25}\smash{\begin{tabular}[t]{l}\textit{4}\end{tabular}}}}%
    \put(0.45973617,0.21640439){\color[rgb]{0,0,0}\makebox(0,0)[lt]{\lineheight{1.25}\smash{\begin{tabular}[t]{l}\textit{5}\end{tabular}}}}%
    \put(0.62153194,0.21640439){\color[rgb]{0,0,0}\makebox(0,0)[lt]{\lineheight{1.25}\smash{\begin{tabular}[t]{l}\textit{6}\end{tabular}}}}%
  \end{picture}%
\endgroup%

%% file: per_disl_mod.pdf_tex
%% Accompanies image file 'per_disl_mod.pdf' (pdf, eps, ps)
\begingroup%
  \makeatletter%
  \providecommand\color[2][]{%
    \errmessage{(Inkscape) Color is used for the text in Inkscape, but the package 'color.sty' is not loaded}%
    \renewcommand\color[2][]{}%
  }%
  \providecommand\transparent[1]{%
    \errmessage{(Inkscape) Transparency is used (non-zero) for the text in Inkscape, but the package 'transparent.sty' is not loaded}%
    \renewcommand\transparent[1]{}%
  }%
  \providecommand\rotatebox[2]{#2}%
  \newcommand*\fsize{\dimexpr\f@size pt\relax}%
  \newcommand*\lineheight[1]{\fontsize{\fsize}{#1\fsize}\selectfont}%
  \ifx\svgwidth\undefined%
    \setlength{\unitlength}{348.66141732bp}%
    \ifx\svgscale\undefined%
      \relax%
    \else%
      \setlength{\unitlength}{\unitlength * \real{\svgscale}}%
    \fi%
  \else%
    \setlength{\unitlength}{\svgwidth}%
  \fi%
  \global\let\svgwidth\undefined%
  \global\let\svgscale\undefined%
  \makeatother%
  \begin{picture}(1,0.71544715)%
    \lineheight{1}%
    \setlength\tabcolsep{0pt}%
    \put(0,0){\includegraphics[width=\unitlength,page=1]{per_disl_mod.pdf}}%
    \put(0.46913745,0.02167673){\color[rgb]{0,0,0}\makebox(0,0)[lt]{\lineheight{1.25000012}\smash{\begin{tabular}[t]{l}2\end{tabular}}}}%
    \put(0.51614143,0.02199208){\color[rgb]{0,0,0}\makebox(0,0)[lt]{\lineheight{1.25000012}\smash{\begin{tabular}[t]{l}2.1\end{tabular}}}}%
    \put(0.57398921,0.02207817){\color[rgb]{0,0,0}\makebox(0,0)[lt]{\lineheight{1.25000012}\smash{\begin{tabular}[t]{l}2.3\end{tabular}}}}%
    \put(0.81307669,0.02207817){\color[rgb]{0,0,0}\makebox(0,0)[lt]{\lineheight{1.25000012}\smash{\begin{tabular}[t]{l}4.5\end{tabular}}}}%
    \put(0.90167077,0.02207817){\color[rgb]{0,0,0}\makebox(0,0)[lt]{\lineheight{1.25000012}\smash{\begin{tabular}[t]{l}5\end{tabular}}}}%
    \put(0.95561946,0.02207817){\color[rgb]{0,0,0}\makebox(0,0)[lt]{\lineheight{1.25000012}\smash{\begin{tabular}[t]{l}5.2\end{tabular}}}}%
    \put(0.74785796,0.02167673){\color[rgb]{0,0,0}\makebox(0,0)[lt]{\lineheight{1.25}\smash{\begin{tabular}[t]{l}4\end{tabular}}}}%
    \put(0.74002648,0.34810007){\color[rgb]{0,0,0}\makebox(0,0)[lt]{\lineheight{1.25}\smash{\begin{tabular}[t]{l}3.5\end{tabular}}}}%
    \put(0.66720462,0.34810007){\color[rgb]{0,0,0}\makebox(0,0)[lt]{\lineheight{1.25}\smash{\begin{tabular}[t]{l}3.3\end{tabular}}}}%
    \put(0,0){\includegraphics[width=\unitlength,page=2]{per_disl_mod.pdf}}%
    \put(0.63327575,0.02230767){\color[rgb]{0,0,0}\makebox(0,0)[lt]{\lineheight{1.25000012}\smash{\begin{tabular}[t]{l}2.95\end{tabular}}}}%
    \put(0,0){\includegraphics[width=\unitlength,page=3]{per_disl_mod.pdf}}%
    \put(0.58730104,0.34810007){\color[rgb]{0,0,0}\makebox(0,0)[lt]{\lineheight{1.25}\smash{\begin{tabular}[t]{l}3.05\end{tabular}}}}%
    \put(0,0){\includegraphics[width=\unitlength,page=4]{per_disl_mod.pdf}}%
    \put(0.13050415,0.34923886){\color[rgb]{0,0,0}\makebox(0,0)[lt]{\lineheight{1.25000012}\smash{\begin{tabular}[t]{l}0\end{tabular}}}}%
    \put(0.25395586,0.34903814){\color[rgb]{0,0,0}\makebox(0,0)[lt]{\lineheight{1.25000012}\smash{\begin{tabular}[t]{l}1\end{tabular}}}}%
    \put(0.35013309,0.34906692){\color[rgb]{0,0,0}\makebox(0,0)[lt]{\lineheight{1.25000012}\smash{\begin{tabular}[t]{l}1.5\end{tabular}}}}%
    \put(0.43136388,0.34919594){\color[rgb]{0,0,0}\makebox(0,0)[lt]{\lineheight{1.25000012}\smash{\begin{tabular}[t]{l}1.7\end{tabular}}}}%
    \put(0.49829798,0.34906692){\color[rgb]{0,0,0}\makebox(0,0)[lt]{\lineheight{1.25000012}\smash{\begin{tabular}[t]{l}1.85\end{tabular}}}}%
    \put(0.50258434,0.67445626){\color[rgb]{0,0,0}\makebox(0,0)[lt]{\lineheight{1.25000012}\smash{\begin{tabular}[t]{l}2\end{tabular}}}}%
    \put(0.5684453,0.67485783){\color[rgb]{0,0,0}\makebox(0,0)[lt]{\lineheight{1.25000012}\smash{\begin{tabular}[t]{l}2.3 - 6\end{tabular}}}}%
    \put(0.01993884,0.50976376){\color[rgb]{0,0,0}\makebox(0,0)[lt]{\lineheight{1.25}\smash{\begin{tabular}[t]{l}a)\end{tabular}}}}%
    \put(0.02101151,0.18817288){\color[rgb]{0,0,0}\makebox(0,0)[lt]{\lineheight{1.25}\smash{\begin{tabular}[t]{l}b)\end{tabular}}}}%
  \end{picture}%
\endgroup%